\documentclass[a4paper,nofootinbib,twocolumn]{revtex4-1}

\usepackage{epsfig}
\usepackage{amsmath,amsfonts,amssymb}
\usepackage{t1enc}

\providecommand{\openone}{\leavevmode\hbox{\small1\kern-3.8pt\normalsize1}}

\newcommand{\Vl}{V_L}
\newcommand{\Vr}{V_R}
\newcommand{\gl}{g_L}
\newcommand{\gr}{g_R}

\newcommand{\fz}{F_0}
\newcommand{\fl}{F_L}
\newcommand{\fr}{F_R}

\begin{document}

\title{Studying the $Wtb$ vertex structure using recent LHC results}

\author{
C\'esar Bernardo$^1$, 
N. F. Castro$^2$, 
Miguel C. N. Fiolhais$^{3,4}$, 
Hugo Gon\c{c}alves$^1$, 
Andr\'e G. C. Guerra$^5$, 
Miguel Oliveira$^{5}$, 
A. Onofre$^{2}$
\\[3mm]
{\footnotesize {\it 
$^1$ Centro de F\'{\i}sica, Universidade do Minho, Campus de Gualtar, 4710-057 Braga, Portugal\\ 
$^2$ LIP, Departamento de F\'{\i}sica, Universidade do Minho, 4710-057 Braga, Portugal\\
$^3$ LIP, Departamento de F\'{\i}sica, Universidade de Coimbra, 3004-516 Coimbra, Portugal\\
$^4$ Department of Physics, City College of the City University of New York, \\ 
     160 Convent Avenue, New York 10031, NY, USA \\
$^5$ Departamento de F\'{\i}sica e Astronomia, Faculdade de Ci\^encias, \\[-1mm] 
     Universidade do Porto, Rua do Campo Alegre 687, 4169-007 Porto, Portugal
}}
}

\begin{abstract}
The $Wtb$ vertex structure and the search for new anomalous couplings is studied using top quark measurements obtained at the LHC, for a 
centre-of-mass energy of 8~TeV. By combining the latest and most precise results on the single top quark production cross section and the 
measurements of the $W$-boson helicity fractions ($\fz$ and $\fl$), it is possible to set new limits, at 95\%~CL (confidence level), on the 
real and imaginary components of the new couplings. The combination of the LHC observables clearly improves the limits obtained when using the 
individual results alone. The updated measurements of the $W$-boson helicity fractions and the $s+t$-channels electroweak single top 
quark production, at the Tevatron, improve the LHC limits, when a world combination of all observables (LHC+Tevatron) is performed.
\end{abstract}

%\pacs{12.15.Ff,12.15.Hh,12.60.-i,14.65.Ha}
\keywords{top quark; anomalous couplings; LHC collider}

\maketitle

\section{Introduction}

The discovery of the top quark at the Tevatron in 1995~\cite{top-CDF, top-D0} was the starting point of a new era of precision studies for 
the Standard Model (SM). With the amount of data already collected by the LHC experiments (ATLAS and CMS), precision tests of the top quark 
properties have become more and more common. Nonetheless, the possibility of existence of physics, beyond the SM, hidden in the uncertainties 
of the experimental measurements, requires a criterious survey of the parameter space in the quest for signs of new contributions in the top 
quark sector. In the SM, the top quark decays almost exclusively to a $W$-boson and a $b$-quark and the decay vertex has the typical (V-A) 
structure. One way to probe the structure of the vertex and test the SM is to study the helicity fractions of the $W$-bosons produced in top 
quark decays, and/or the observables which depend on these helicity fractions~\cite{delAguila:2002nf, AguilarSaavedra:2006fy} {\it i.e.}, 
angular and spin asymmetries, ratios of fractions, etc. The Tevatron results on the helicity fractions, obtained by CDF and D0, have been 
combined~\cite{whel-CDF_D0}, showing good agreement (within 10\%) with the NNLO SM predictions~\cite{Czarnecki:2010gb}. At the LHC, both 
ATLAS~\cite{Wh-ATLAS} and CMS~\cite{Wh-CMS} have measured the $W$-boson helicity fractions in top quark decays. The production cross sections 
can play an important role, as well, in constraining the allowed phase space for the new couplings. The $t$-channel single top quark 
production is particularly relevant given its magnitude. Measurements of the $t$-channel cross section have been performed by both the ATLAS 
~\cite{st-ATLAS} and CMS~\cite{st-CMS} experiments at centre-of-mass energies of 7 and 8~TeV. Although smaller then the $t$-channel, the
 $Wt$ associated production cross section is also important. In fact, as the $Wt$ cross section is not affected by 
new physics contributions like, for instance, four-fermion contact interactions, it can probe the $Wtb$ vertex in a model independent way. 
ATLAS~\cite{Wt-ATLAS} and CMS~\cite{Wt-CMS} measured the $Wt$ associated production at 7 and 8~TeV. 

In this paper, the results of the most precise measurements of the single top quark production cross section at a centre-of-mass energy 
of 8~TeV at the LHC, are combined with the $W$-boson helicity fractions ($\fz$ and $\fl$), obtained by the CMS experiment 
to set limits on the real and imaginary parts of possible new anomalous couplings that could be present at the $Wtb$ vertex. 
The {\sc TopFit}~\cite{topfit} code is used to perform the combination of the observables, as well as to extract the corresponding 
anomalous couplings allowed regions at 95\% confidence level (CL). The real and imaginary part of the couplings are 
considered. In order to describe the interaction, an effective field theory approach is used~\cite{Buchmuller:1985jz}. 
The most general vertex can be described by~\cite{AguilarSaavedra:2008zc}

\begin{eqnarray}
\mathcal{L}_{Wtb} & = & - \frac{g}{\sqrt 2} \bar b \, \gamma^{\mu} \left( \Vl
P_L + \Vr P_R
\right) t\; W_\mu^- \nonumber \\
& & - \frac{g}{\sqrt 2} \bar b \, \frac{i \sigma^{\mu \nu} q_\nu}{M_W}
\left( \gl P_L + \gr P_R \right) t\; W_\mu^- + \mathrm{H.c.} \,, \notag \\
\label{ec:lagr}
\end{eqnarray}
where $\Vl$, $\Vr$, $\gl$, $\gr$ are dimensionless couplings that are, in general, complex. At tree level and in the SM, $\Vl = 
V_{tb} \simeq 1$ and all the other couplings vanish. However, if new physics appear at the $Wtb$ vertex, the new 
couplings $\Vr$, $\gl$ and $\gr$ may acquire important corrections. These anomalous couplings can 
be tested in top quark decays by measuring the $W$-boson helicity states~\cite{Kane:1991bg, delAguila:2002nf, 
AguilarSaavedra:2006fy}. The results can be combined with the single top quark production measurements
to improve the limits on the anomalous couplings~\cite{Boos:1999dd, Najafabadi:2008pb, AguilarSaavedra:2008gt,Chen:2005vr, 
AguilarSaavedra:2010nx, Zhang:2010dr}. It is interesting to recall the sensitivity to the new couplings at the LHC, in top quark decays,
can be largely surpassed at a future Linear Collider (ILC)~\cite{AguilarSaavedra:2012vh}.

\section{LHC observables at 8~TeV}

\subsection{$W$-boson helicity fractions}

The $W$-boson helicity fractions in this paper were measured by the CMS experiment at a centre-of-mass energy of 
8~TeV using a sample of $t \bar t$ events which decayed to the muon+jets channel. The sample corresponds to an 
integrated luminosity of 19.6~fb$^{-1}$. The longitudinal and left handed helicity fractions are, 
respectively~\cite{Wh-CMS},
\begin{align}
& \fz = 0.659 \pm 0.015\;\text{(stat)} \pm 0.023\;\text{(syst)} \,, \notag \\
& \fl = 0.350 \pm 0.010\;\text{(stat)} \pm 0.024\;\text{(syst)} \,, 
\end{align}
with a correlation coefficient $\rho = -0.95$ and assuming a top quark mass $m_t = 172.5$ GeV. The measurement is 
clearly dominated by the systematic uncertainties.

\subsection{Single top quark production cross sections}

The most precise measurement of the $t$-channel single top quark production cross section used in this paper, was performed 
by the CMS experiment~\cite{st-CMS} at 8~TeV. Signal events were considered whenever top-quark decay products were accompained with 
a high rapidity light quark together with a low $p_T$ $b$-quark. Events were selected if they contained one isolated 
lepton (electron or muon). The value of the top quark mass assumed in all simulated Monte Carlo samples 
was $m_t = 172.5$ GeV. The measurement, using a total integrated luminosity of 19.7~fb$^{-1}$ is,
\begin{align}
& \sigma_{t-chan} = 83.6 \pm 2.3\;\text{(stat)} \pm 7.4\;\text{(syst)} ~{\rm pb}\,.
\end{align}

The effect of the $Wt$ associated production in the anomalous couplings fits, is also investigated. The measurement with
smaller uncertainty, normalized to the SM cross section expectation, is~\cite{Wt-CMS},
\begin{align}
& \sigma_{Wt} = 23.4 \pm 5.4 ~{\rm pb} \,.
\end{align}
The measurements of the single top quark production cross sections are assumed to be uncorrelated.

\section{95\%~CL Limits at the LHC}

By taking into account the analytic expressions introduced in~\cite{AguilarSaavedra:2006fy} 
and~\cite{AguilarSaavedra:2008gt} for the helicity fractions and single top quark production cross section as 
a function of the complex anomalous couplings, respectivelly, it is possible to determine, using {\sc TopFit}, the allowed regions 
for the couplings, provided a minimal set of measurements is specified. 
In the current paper, the CMS results of the helicity fractions ($\fz$ and $\fl$) and the $t$-channel single top quark production cross section are, 
initially, used as input parameters to {\sc TopFit}. It should be stressed that no four-fermion contributions to the $t$-channel single top quark 
production cross section is considered~\cite{Cao:2007ea, AguilarSaavedra:2010zi}. No correlations are assumed between the helicities and the cross 
section measurements. The total uncertainty associated to each measurement is defined by adding in quadrature the corresponding statistical and 
systematic uncertainties. The results are presented in terms of two dimensional plots of subsets of anomalous couplings (assuming the others 
vanishing) or individual limits (assuming all other anomalous couplings vanishing), as convenient, to ilustrate each physics case. It should be 
stressed that the approach, although not fully general, is justified once different couplings indeed arise from different gauge invariant operators 
and the limited number of observables and their precision makes a global fit almost useless for the time being. Limits are set by {\sc TopFit}, at 
95\%~CL, for a top quark mass of $m_t = 172.5$ GeV (as for the LHC experiments), $M_W = 80.4$ GeV and $m_b = 4.8$ GeV.
Two different scenarios are considered:
\begin{itemize}
\item the couplings are assumed real ($Re$) and,
\item the couplings may have a non vanishing imaginary ($Im$) part.
\end{itemize}
In both scenarios several combinations of couplings are considered in order to ilustrate the potential of the physics case and give an idea of the 
order of magnitude of the 95\%~CL allowed regions for the anomalous couplings, given the current results at the LHC. The effect of the $Wt$ associated 
production cross section at the LHC, as well as the recent results on the $W$-boson helicity fractions and the $s+t$ single top quark production cross 
section measured at the Tevatron, are finally combined with the previous measurements to set 95\%~CL limits on the anomalous couplings allowed 
regions.
\subsection{Real anomalous couplings}
\begin{figure*}[h]
\begin{center}
\begin{tabular}{ccc}
\epsfig{file=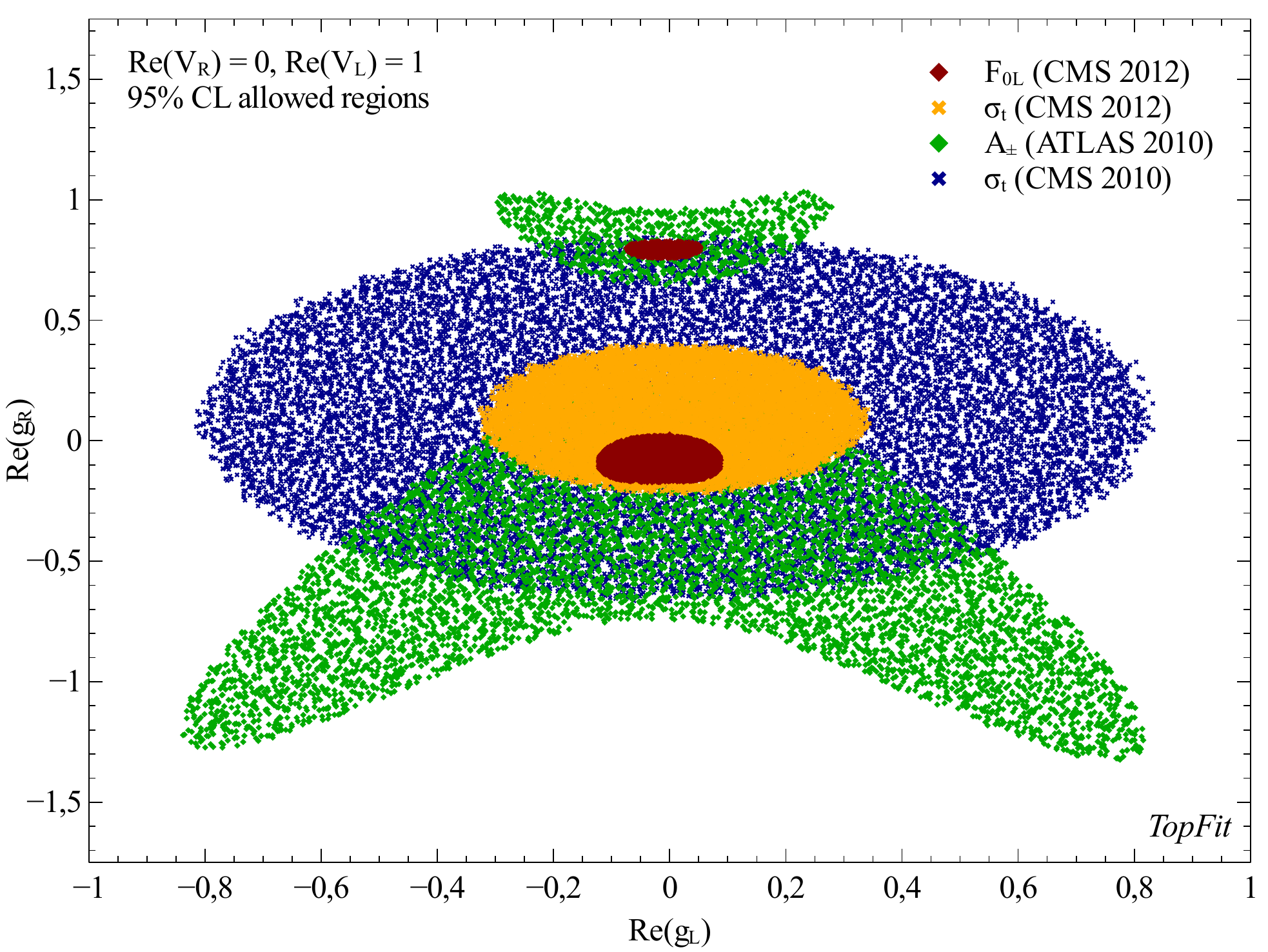,height=5.5cm,clip=} & \quad &
\epsfig{file=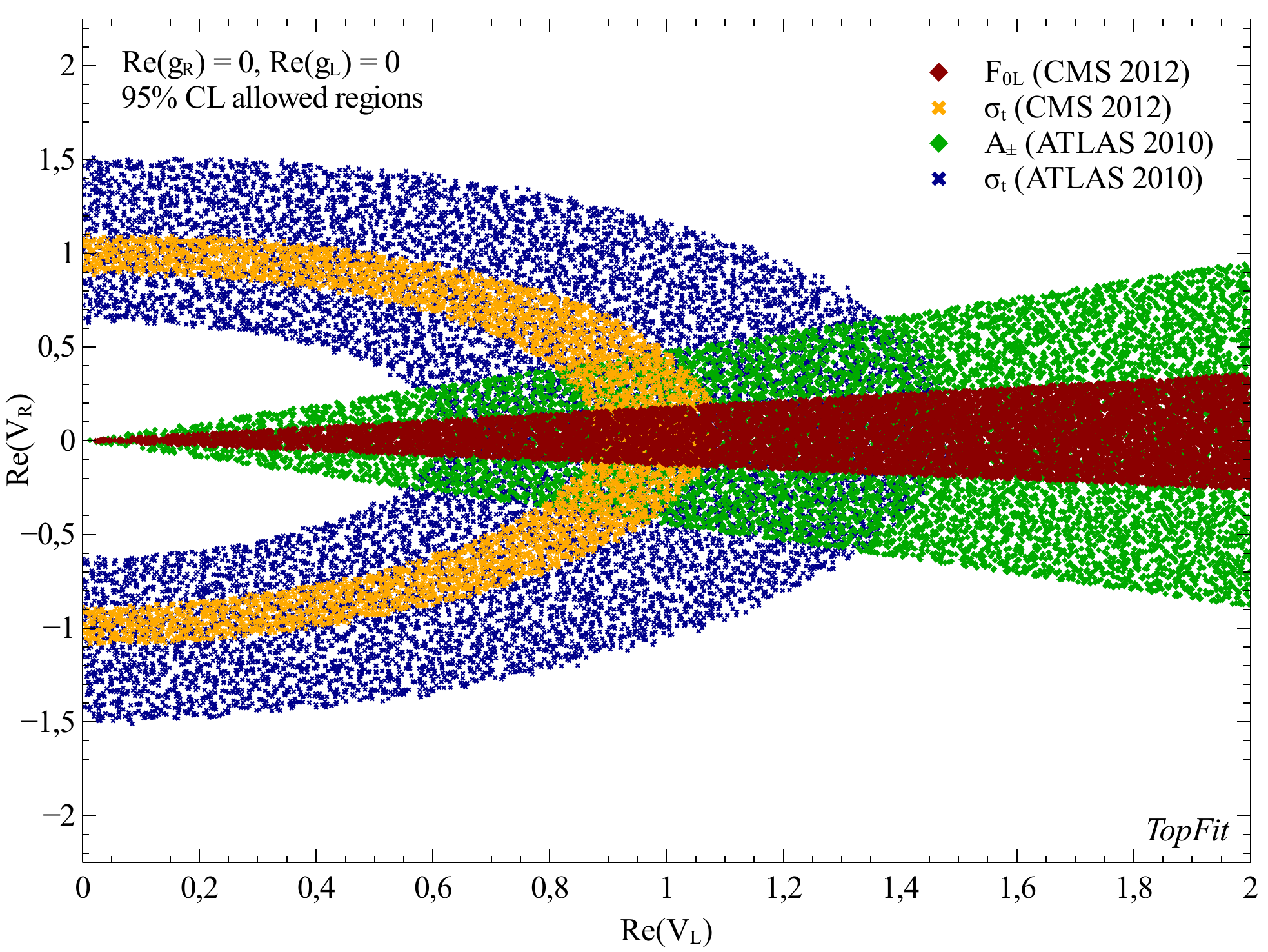,height=5.5cm,clip=}
\end{tabular}
\caption{LHC 95\%~CL allowed regions in the planes [$Re(\gr),Re(\gl)$] (left) and [$Re(\Vr),Re(\Vl)$] (right) from 
$W$-boson helicity and $t$-channel single top quark production at the LHC assuming couplings are real.}
\label{fig:lim1}
\end{center}
\end{figure*}

In the following discussion, all couplings are assumed real. The $W$-boson helicity fractions ($\fz$ and $\fl$) and the $t$-channel single top quark 
production cross section at the LHC were used as input parameters to {\sc TopFit}. In Fig.~\ref{fig:lim1} the limits obtained with the 2012 
CMS~\cite{Wh-CMS, st-CMS} results are compared with the previous results obtained in 2010~\cite{AguilarSaavedra:2011ct}. In the left plot, limits at 
95\%~CL are set in the $[Re(\gr),Re(\gl)]$ plane, assuming $\Vr$ = 0 and $\Vl$=1. In the right plot, limits in the plane $[Re(\Vr),Re(\Vl)]$ are 
shown, for $\gr = \gl = 0$.
\begin{table}[h]
\begin{center}
\begin{tabular}{|c|c|c|c|}
\hline
         LHC                       &      $\gr$      &       $\gl$       &     $\Vr$      \\ \hline
 \hline
         Allowed Regions ($Re$)    & [-0.15 , 0.01]  &   [-0.09 , 0.06]  & [-0.13 , 0.18] \\
\hline
\end{tabular}
\caption{One dimension 95\%~CL limits on the anomalous couplings (assumed real) from $W$-boson helicities and $t$-channel cross section at the LHC.} 
\label{tab:ReCoup}
\end{center}
\end{table}
A clear improvement with respect to the 2010 results can be observed. The complementarity of the different measurements is clearly visible, 
as discussed in~\cite{AguilarSaavedra:2011ct}: when allowed regions corresponding to different measurements overlap, either for 2010 or 2012, 
the combined measurements allowed region is indeed very much constrained, when compared to the results from the individual measurements alone. 
In Table~\ref{tab:ReCoup} the 95\%~CL limits for the couplings, assuming $\Vl=1$ and all the other (real) couplings zero at a time, are shown.
\subsection{Complex anomalous couplings}
\begin{figure*}[h]
\begin{center}
\begin{tabular}{ccc}
\epsfig{file=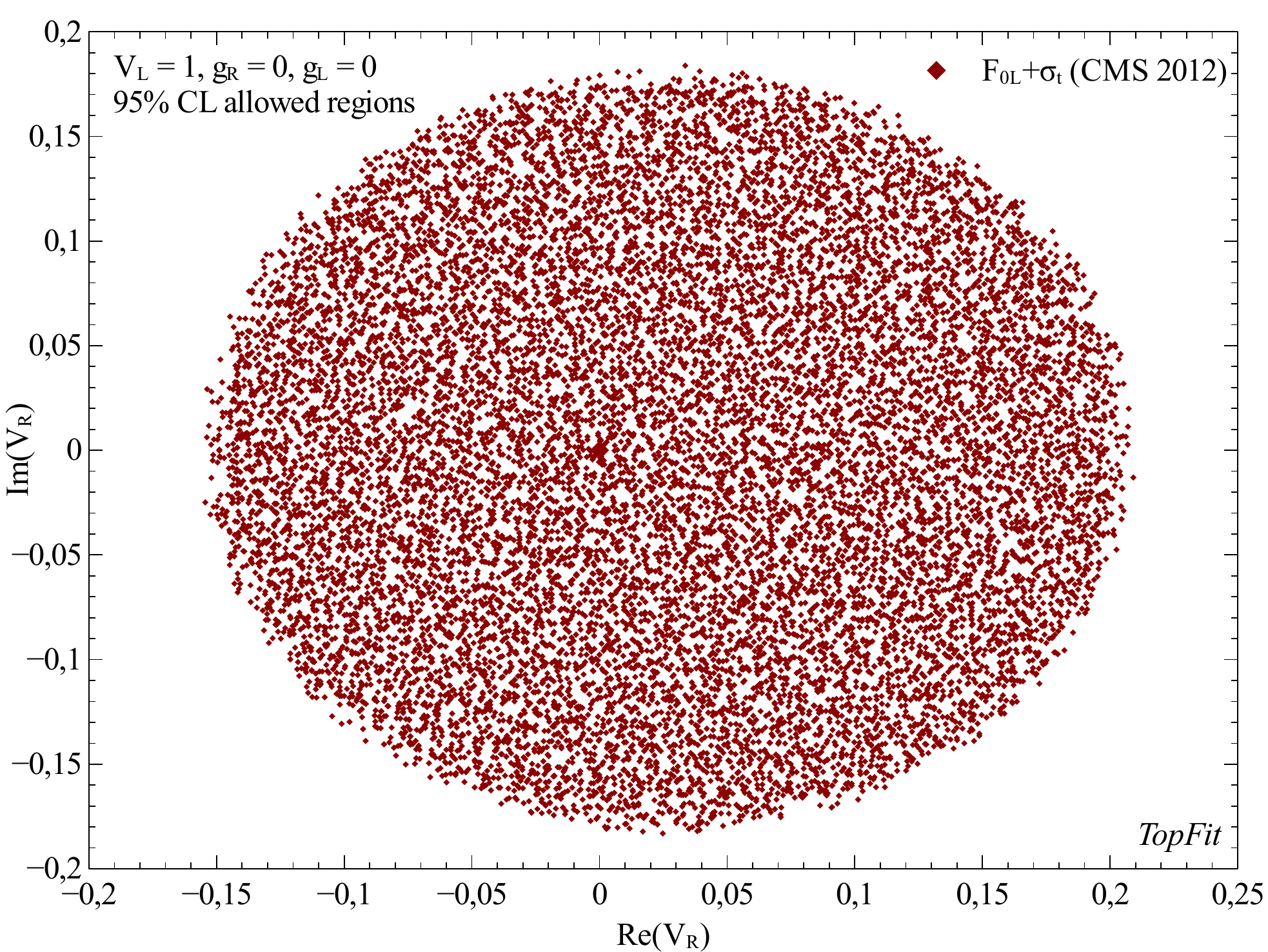,height=5.5cm,clip=} & \quad &
\epsfig{file=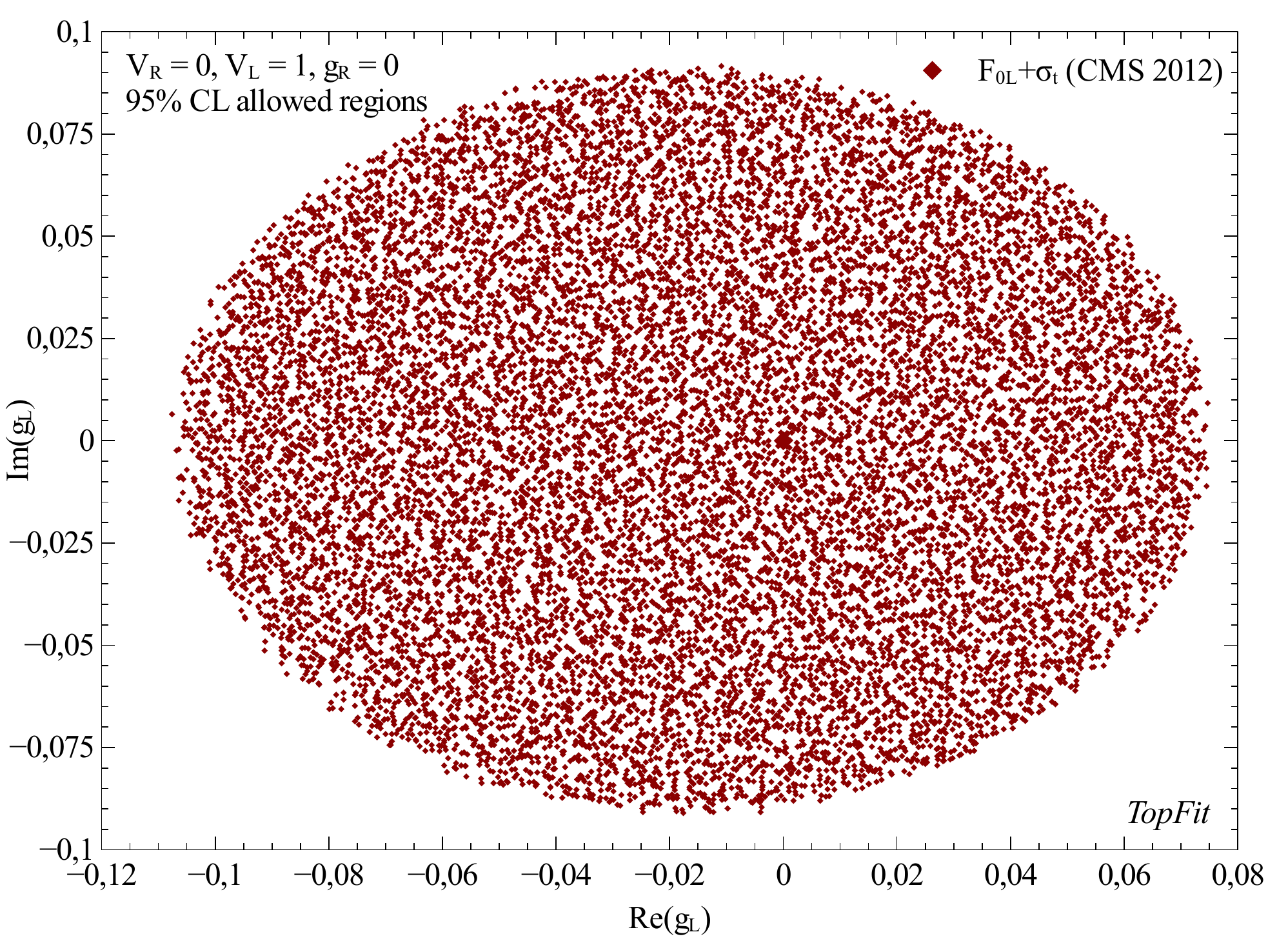,height=5.5cm,clip=}
\end{tabular}
\caption{95\%~CL allowed regions for the $\Vr$ (left) and $\gl$ (right) complex couplings from the combination of the 
$W$-boson helicity and $t$-channel single top quark production measurements at the LHC.}
\label{fig:lim2}
\end{center}
\end{figure*}

\begin{figure*}[h]
\begin{center}
\begin{tabular}{ccc}
\epsfig{file=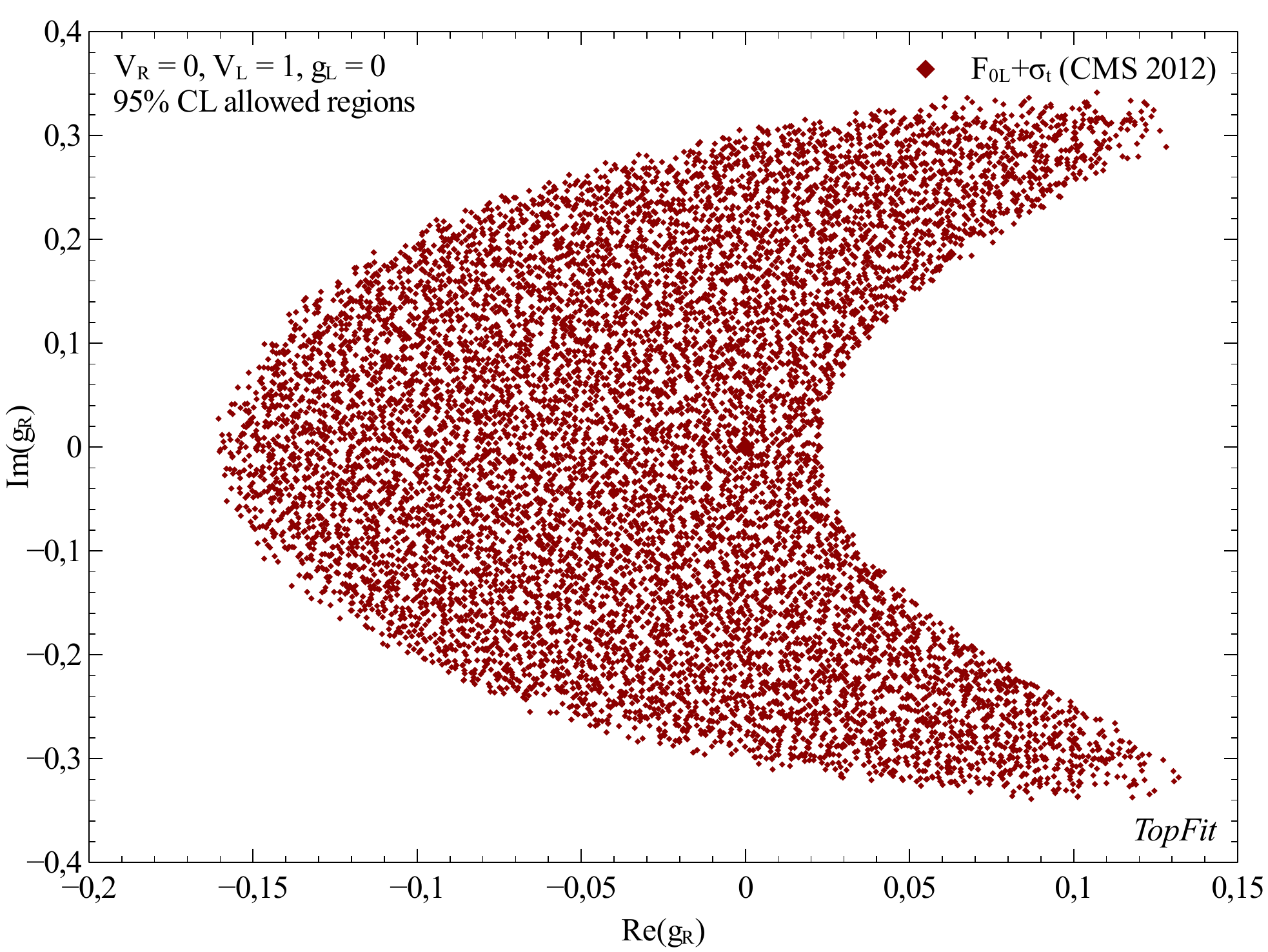,height=5.5cm,clip=} & \quad &
\epsfig{file=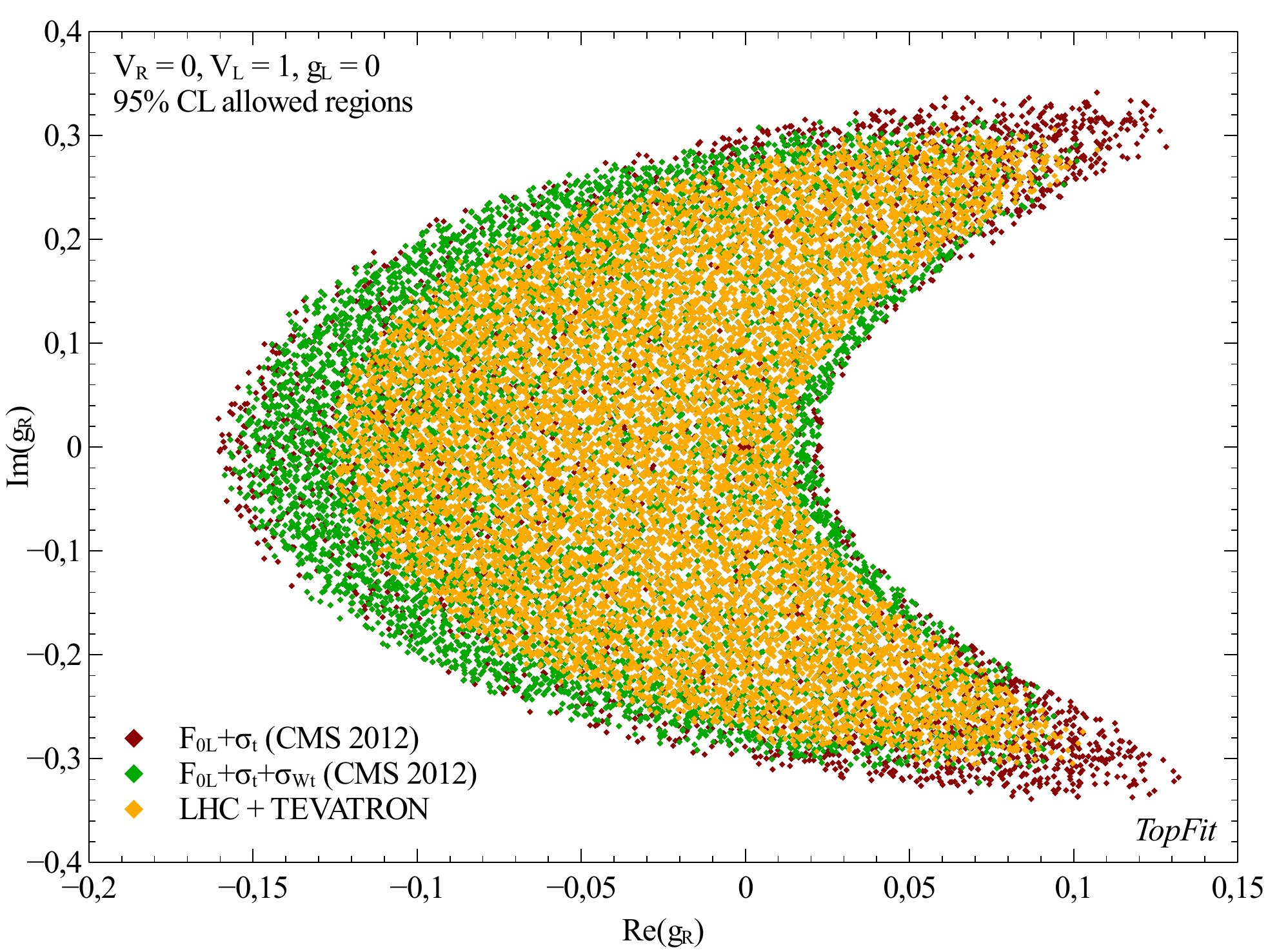,height=5.5cm,clip=}
\end{tabular}
\caption{95\%~CL allowed regions for the $\gr$ complex coupling from the combination of the $W$-boson helicity and $t$-channel single top 
production measurements at the LHC (left) and combining all LHC and Tevatron results (right).}
\label{fig:lim3}
\end{center}
\end{figure*}

In this section the anomalous couplings $\gr$, $\gl$ and $\Vr$ are assumed complex, with both real and imaginary components. By combining the helicity 
fractions with the $t$-channel single top quark production cross section results at the LHC, limits at 95\%~CL were set on the allowed regions of the 
$Im$ versus the $Re$ components of the couplings.
\begin{table}[h]
\begin{center}
\begin{tabular}{|c|c|c|c|}
\hline
               LHC            &        $\gr$      &       $\gl$        &     $\Vr$ \\ 
\hline
     Allowed Regions ($Re$)   &   [-0.16 , 0.13]  &    [-0.11 , 0.08]  & [-0.15 , 0.21] \\
     Allowed Regions ($Im$)   &   [-0.34 , 0.34]  &    [-0.09 , 0.09]  & [-0.18 , 0.18] \\
\hline
\hline
           LHC+Tevatron       &        $\gr$      &       $\gl$        &     $\Vr$ \\
\hline
     Allowed Regions ($Re$)   &   [-0.13 , 0.11]  &    [-0.10 , 0.07]  & [-0.15 , 0.20] \\
     Allowed Regions ($Im$)   &   [-0.31 , 0.31]  &    [-0.09 , 0.09]  & [-0.17 , 0.17] \\
\hline
\end{tabular}
\caption{Two dimension 95\%~CL limits on the real and imaginary components of the anomalous couplings from $W$-boson helicities and $t$-channel 
cross section at the LHC (top), and from the combination of the LHC and Tevatron measurements (bottom).}
\label{tab:ImReCoup}
\end{center}
\end{table}
\begin{table}[h]
\begin{center}
\begin{tabular}{|c|c|c|c|}
\hline
               LHC            &        $\gr$    &       $\gl$        &     $\Vr$ \\ 
\hline
     Allowed Regions ($Im$)   & [-0.29 , 0.29]  &    [-0.08 , 0.08]  & [-0.16 , 0.16] \\
\hline
\hline
           LHC+Tevatron       &        $\gr$    &       $\gl$        &     $\Vr$ \\ 
\hline
     Allowed Regions ($Im$)   & [-0.27 , 0.27]  &    [-0.07 , 0.07]  & [-0.15 , 0.15] \\
\hline
\end{tabular}
\caption{One dimension 95\%~CL limits on pure imaginary anomalous couplings from $W$-boson helicities and $t$-channel 
cross section at the LHC (top) and from the combination of the LHC and Tevatron measurements (bottom).} 
\label{tab:ImCoup}
\end{center}
\end{table}
In Fig.~\ref{fig:lim2} the 95\%~CL allowed region for the $\Vr$ (left) and $\gl$ (right) complex couplings is shown, assuming both the $Re$ and the $Im$ 
parts of the couplings non-vanishing. For the left (right) plot the allowed region is obtained by fixing $\gr = \gl = 0, \Vl =1$ ($\Vr = \gr = 0, \Vl 
=1$). In Fig.~\ref{fig:lim3} (left) the 95\%~CL allowed region for the $\gr$ assuming $\Vr = \gl = 0, \Vl =1$, is shown. 
The (two dimensional) LHC limits on the anomalous couplings from Fig.~\ref{fig:lim2} and Fig.~\ref{fig:lim3} (left) are shown in 
Table~\ref{tab:ImReCoup} (top). One particular case is considered when the anomalous couplings are pure imaginary only.  In 
Table~\ref{tab:ImCoup} (top), the 95\%~CL limits for the {\it Im} part of the couplings, assuming $\Vl=1$ and all the other 
couplings vanishing at a time (as well as any {\it Re} part), are shown.
\subsection{Impact of $Wt$ associated production at LHC and measurements at the Tevatron}

The effect of the LHC $Wt$ associated production cross section was studied by including in the fit, as well, its most precise measurement at 8~TeV 
($\sigma_{Wt} = 23.4 \pm 5.4$~pb~\cite{Wt-CMS}). The improvement seems limited, as can be noticed in Fig.~\ref{fig:lim3} (right, green dots or lighter 
gray dots if printed in black and white). A more visible effect is observed, when the $W$-boson helicity fractions and the single top quark cross 
section measurements at the Tevatron are included, in addition to the LHC results. The combined Tevatron measurements of the $W$-boson helicity 
fractions ($\fz=$0.722$\pm$0.081, $\fr=$-0.033$\pm$0.046, with correlation -0.88 ~\cite{Wh-CDF-D0}) and the ($s+t$) single top quark cross section 
with smaller uncertainty at the Tevatron ($\sigma_{s+t}$=3.02$^{+0.49}_{-0.48}$~pb~\cite{stXsec-CDF}), were used. The (two dimensional) LHC limits on 
the anomalous couplings from Fig.~\ref{fig:lim3} (right, yellow dots or darker dots if printed in black and white) are shown in 
Table~\ref{tab:ImReCoup} (bottom). In Table~\ref{tab:ImCoup} (bottom), the 95\%~CL limits for the {\it Im} part of the couplings, assuming $\Vl=1$ and 
all the other couplings vanishing at a time (as well as any {\it Re} part), are shown. The Tevatron measurements can improve the LHC limits by as much 
as roughly 20\%, specially for the $\gr$ anomalous coupling.

\section{Conclusions}

In this paper, 95\%~CL limits on new anomalous couplings at the $Wtb$ vertex are revisited. Recent measurements at 8~TeV, of the $W$-boson helicity fractions 
($\fz$ and $\fl$) and single top quark production cross sections ($t$-channel and $Wt$ associated production) at the LHC, were used. It was 
shown, with simple case studies, that a proper combination of the LHC observables can provide useful information on anomalous couplings. This is true 
not only for the $Re$ part of the couplings, but also for the $Im$ part. Similar 95\%~CL limits for the {\it Re} and {\it Im} components of the anomalous 
couplings are observed. This suggests that, the data from the LHC can be used to constrain equally well, not only the real, but also the imaginary part of 
the couplings. Given the current precision of the measurements at the LHC and the Tevatron, it is still useful to combine the results from both 
colliders. Improvements as high as roughly 20\% in the limits, in particular for the anomalous coupling $\gr$, are observed.

\vspace{0.5cm}

\section*{Acknowledgements}

This work was partially supported by Funda\c{c}\~ao para a Ci\^encia e Tecnologia, FCT (project CERN/FP/123619/2011, grant SFRH/BI/52524/2014 and contract 
IF/00050/2013). The work of M.C.N.~Fiolhais was supported by LIP-Laborat\'orio de Instrumenta\c c\~ao e F\'isica Experimental de Part\'iculas, 
Portugal (grant PestIC/FIS/LA007/2013). The authors would like to thank the support of CRUP (Conselho de Reitores das Universidades Portuguesas) 
through Ac\c{c}\~ao integrada Ref. E 2/09 and the MAP-Fis Program (the Joint Doctoral Programs of the Universities of Minho, Aveiro and Porto, 
http://www.map.edu.pt/fis/home). Special thanks go to Juan Antonio Aguilar-Saavedra for all the fruitful discussions and a long term collaboration.

% Figures

\end{document}